# Frustration of Exchange-Striction Ferroelectric Ordering and Growth of Electric Polarization of Phase Separation Domains in $R_{0.8}Ce_{0.2}Mn_2O_5$ (R = Er, Tb) Solid Solutions

B. Kh. Khannanov [a], E. I. Golovenchits [a], and V. A. Sanina [a], *

[a] Ioffe Institute, Russian Academy of Sciences, St. Petersburg, 194021 Russia

The effect of rare-earth ions (R = $Er^{3+}$, $Tb^{3+}$, and $Ce^{3.75+}$) on the dielectric properties and the electric polarization induced by local polar phase separation domains in solid solutions of $R_{0.8}Ce_{0.2}Mn_2O_5$ (R = Er, Tb) multiferroics has been studied. These parameters are found to qualitatively differ from those of initial $RMn_2O_5$ (R = Er, Tb) crystals studied before. It is shown that the properties of the polar phase separation domains that form in a subsystem of $Mn^{3+}$ and $Mn^{4+}$ ions due to a finite probability of tunneling electrons between these ions with different valences are substantially dependent on the values of crystal fields in which these domains exist. A combined influence of $Er^{3+}$, $Tb^{3+}$, and $Ce^{3.75+}$ ions is found to substantially change the crystal field in $R_{0.8}Ce_{0.2}Mn_2O_5$ (R = Er, Tb) as compared to $RMn_2O_5$ (R = Er, Tb).

## 1. INTRODUCTION

In [1], the results of studies of the electric polarization induced by local polar phase separation domains in $RMn_2O_5$ (R = Er, Tb) are presented and compared to the properties of $RMn_2O_5$ (R = Gd, Bi) [2–4]. All these crystals are multiferroics of the second type, in which, at temperatures lower than the Curie temperature ($T = T_C \leq 30$–35 K), there is a ferroelectric ordering induced by magnetic ordering with the Neel temperature ($T_N \leq 35$–40 K) [5, 6]. In [1–4], it was shown that local polar phase separation domains are formed in these multiferroics over a wide temperature range from the lowest temperatures to those significantly higher than $T_C$. Below some temperatures which are dependent on the crystal axis directions and are significantly higher than $T_C$ of the low-temperature ferroelectric ordering, these domains form the frozen superparaelectric state, in which the response of the local polar domains on an applied electric field has the shape of the hysteresis loops of electric polarization with the remanent polarization.

A specific feature of $RMn_2O_5$ is the same numbers of manganese ions $Mn^{3+}$ (containing three $t_{2g}$ and one $e_g$ electrons on 3d shell) and $Mn^{4+}$ (with three $t_{2g}$ electrons on the 3d shell), which provides the conditions for the appearance of the dielectric charge ordering. The $Mn^{4+}$ ions have the octahedral oxygen environment and are located in layers with z = 0.25c and (1 − z) = 0.75c. The $Mn^{3+}$ ions have a noncentral local environment in the shape of pentagonal pyramids and are disposed in layers with z = 0.5c. The $R^{3+}$ ions with the environment similar to that of $Mn^{3+}$ are disposed in layers with z = 0 [7]. The charge ordering and the finite probability of $e_g$ electrons transfer between $Mn^{4+}$–$Mn^{3+}$ ion pairs (double exchange [8, 9]) are key factors which determine the electric polar state of $RMn_2O_5$ at all temperatures. The low-temperature ferroelectric state at $T \leq T_C$ is due to the charge ordering along the axis *b*. The alternating of neighboring $Mn^{3+}$ and $Mn^{4+}$ ion pairs along the axis b with a strong ferromagnetic exchange (double exchange) and a weak indirect antiferromagnetic exchange leads to the exchange striction that distorts the centrosymmetricity of the lattice along axis *b*, and the appearance of the low-temperature ferroelectric ordering [10]. On the other hand, the transport of $e_g$ electrons between $Mn^{3+}$–$Mn^{4+}$ ion pairs disposed in the neighboring layers perpendicular to the axis *c* leads to the formation of local polar phase separation domains with a different distribution of $Mn^{3+}$ and $Mn^{4+}$ ions as compared to the initial crystal matrix [1–4, 11, 12].

The local polar domains are formed in $RMn_2O_5$ due to the phase separation processes, by analogy with manganites $LaAMnO_3$ (A = Sr, Ca, Ba), which also contain $Mn^{3+}$ and $Mn^{4+}$ ions [9, 13]. The restricted phase separation domains are formed in the initial crystal matrix as a result of self-organization processes due to a finite probability of tunneling $e_g$ electrons between ion pairs ($Mn^{3+}$–$Mn^{4+}$) in the neighboring planes perpendicular to the axis *c*. They are formed due to the balance of strong interactions (the double exchange with a characteristic energy of 0.3 eV, the Jahn–Teller interaction (0.7 eV), and the Coulomb repulsion (1 eV) and exist in a wide temperature range from low temperatures to temperatures higher than room temperature [8, 9, 11, 13]. At low temperatures T < 60 K, the phase separation domains in $RMn_2O_5$ are the restricted 1D superlattices consisting from ferromagnetic layers containing $Mn^{3+}$ and $Mn^{4+}$ ions in various proportions. They demonstrate a set of ferromagnetic resonances [14–16] and the electric polarization [1–4] in the directions of magnetic and electric fields, respectively. At temperatures higher than 180 K, the interaction appears between the local polar domains restricted before that forms 2D superstructures perpendicular to the axis *c*. In such structures, the layers of the initial matrix and the phase separation domains alternate. At room temperature, the layer widths of layers were 700–900 Å [2–4, 11, 12].

The frozen superparaelectric state of local ferroelectric domains in the dielectric centrosymmetric matrix was considered theoretically before [17], but it was observed experimentally for the first time in $RMn_2O_5$ (R = Gd, Bi) [2–4]. In this case, it was shown that the local polar phase separation domains in $RMn_2O_5$ are multiferroics and controlled by external electric and magnetic fields. In those works, it was assumed that the weakly intense non-central reflections observed in the structural studies at room temperature in some $RMn_2O_5$ [18] belonged to these domains.

In [1–4], it was demonstrated that the electric polarization in $RMn_2O_5$ (R = Gd, Bi, Er, and Tb) is due to the phase separation domains that are formed in the Mn ion subsystem but exist in various crystal fields induced by the R ions. The value and the anisotropy of the crystal fields are precisely factors that determine the temperatures of existence and values of electric polarizations along various axes. In this case, the $Gd^{3+}$ ion (the ground state is $^8S_{7/2}$) has the maximum spin among the R ions and it is slightly bounded with the lattice in the *S* state, since there is no spin–orbit interaction in it. The maximum polarization was observed in $GdMn_2O_5$ in the low-temperature ferroelectric ordering [19, 20]. The ground state of $Tb^{3+}$ ions ($^7F_6$, S= 3, L = 3) is characterized by a high magnetic moment (J = 9.7μB), to which both the spin and orbital moments (S = 3, L = 3, respectively) contribute, and there is a strong spin–orbit coupling. The $Tb^{3+}$ ions are described in the extreme strong anisotropic Ising approximation, in which the orientation of their moments is rigidly fixed in plane *ab* [21]. The $Er^{3+}$ ion ground state ($^4I_{15/2}$, S = 3/2, L = 6) also has a high magnetic moment (J = 9.6μB), the main contribution to which is that of the orbital moment and there is a strong electric crystal field that rigidly orients the moments of the Er ions along the axis c by a strong single ion anisotropy [21]. The magnetic moments of the R ions are bounded by the exchange interaction with Mn ions. In addition, R ions strongly change the crystal field in which the Mn ion subsystem exists. The $Bi^{3+}$ ions are nonmagnetic, but they strongly distort the nearest crystal environment due to the existence of lone pairs of $6s^2$ electrons on their outer shells [22].

The aim of this work is to study the $R_{0.8}Ce_{0.2}Mn_2O_5$ (R = Er, Tb) solid solutions and the influence of the dilution of the R-ion subsystem with Ce ions on the low-temperature ferroelectric ordering along axis *b* and also the electric polarization induced by the phase separation domains. Note that these diluted crystals exhibit the same symmetry (space group Pbam) as the initial $RMn_2O_5$ crystals. The crystals containing rare earth Ce ions are characterized by the appearance of the variable valence $Ce^{+3.75}$; i.e., these ions demonstrate the properties of both $Ce^{3+}$ and $Ce^{4+}$ ions. The existence of Ce ions with this variable valence was observed in [23] when studying ceramic $Bi_{0.9}Ce_{0.1}Mn_2O_5$ samples. The Ce ions in single-crystal $R_{0.8}Ce_{0.2}Mn_2O_5$ (R = Er, Tb) studied by us replace of $Er^{3+}$ and $Tb^{3+}$ ion positions. Recall that R ions in $RMn_2O_5$ have the pentagonal oxygen environment (distorted hexahedral). According to [24], in the hexagonal environment the ionic radii of the ions we are interesting are as follows: $Ce^{4+}$ − 0.87 Å, $Ce^{3+}$ − 1.01 Å, $Er^{3+}$ − 0.89 Å, and $Tb^{3+}$ − 0.923 Å. Thus, $Ce^{4+}$ and $Ce^{3+}$ ions can be substituted for both $Er^{3+}$ and Tb3$^+$ ions with some probability.

We studied the influence of Ce ions on the dielectric and magnetic properties of $Eu_{0.8}Ce_{0.2}Mn_2O_5$ before in [11, 14‑16]. It was shown that the dilution of the $Eu^{3+}$ ions mainly with $Ce^{4+}$ ions led to an increase in the concentration of the phase separation domains at all temperatures. This led to an increase in the intensity of the ferromagnetic resonances from the layers of the 1D superlattices that are charged domain walls between the polar domains of ferroelectric ordering [14-16] and also changed the properties of the high-temperature phase separation domains as compared to those of EuMn2O5.

The problem of this work is the study of the influence of dilution of the $RMn_2O_5$ (R = Er, Tb) single crystals with Ce ions on the dielectric properties and the electric polarization in the wide temperature range 5‑350 K.

## 2. EXPERIMENTAL

The $R_{0.8}Ce_{0.2}Mn_2O_5$ (R = Er, Tb) single crystals were grown by spontaneous crystallization from solution–melt [25, 26]. They were 2–3-mm-thick plates with areas of 3–5 mm$^2$. To measure the dielectric properties and the polarization, we fabricated 0.3–0.6-mm-thick flat capacitors with areas of 3–4 mm$^2$. The dielectric permittivity and the conductivity were measured using a Good Will LCR-819 impedance meter at the frequency range 0.5–50 kHz in the temperature range 5–330 K. The electric polarization was measured by the PUND (Positive Up Negative Down) method [27–29]. We used the PUND method adapted to the measurements of the polarization of local polar domains with local conductivity described in [2–4]. The PUND method allows us to take into account the contribution of the conductivity to the measured electric polarization loop. This is particularly important when studying the diluted $R_{0.8}Ce_{0.2}Mn_2O_5$ (R = Er, Tb) with an increased background of the conductivity. In this method, the response of internal polarization P can be distinguished by applying a series of positive P1-P2 and negative N1-N2 voltage pulses to the measured sample. In this case, independent curves (P1-P2) and (N1-N2) of the response of the effective change in polarization to this series of pulses are recorded. The PUND method is based on the difference of the dynamics of the responses of the internal polarization and the conductivity on the pulses of applied electric field E. As the pulse is switched out, the polarization relaxes significantly slower than the conductivity does. The time intervals between the (P1–P2) and (N1–N2) pulses must be such that, in this time, the internal polarization was slightly changed, while the conductivity relaxed completely. The initial internal polarization is calculated by the subtraction of the (P1–P2) and (N1–N2) pair pulses from each other.

## 3. RESULTS AND DISCUSSION

### 3.1. Dielectric Properties and Electric Polarization in $Er_{0.8}Ce_{0.2}Mn_2O_5$

Figures 1a–1f show the temperature dependences of dielectric permittivity $\varepsilon'$ for $Er_{0.8}Ce_{0.2}Mn_2O_5$ ((ECMO) for a number of frequencies along axes *a*, *b*, and *c* (Figs. 1a, 1c, and 1e, respectively) and conductivity $\sigma$ along the same axes (Figs. 1b, 1d, and 1f, respectively). The inset in Fig. 1c shows the values of $\varepsilon'$ along the b axis in a larger scale for some frequencies in $ErMn_2O_5$ (EMO) and ECMO. As it is seen, the dilution of Er ions by Ce ions leads to a strong distortion of the low-temperature ferroelectric ordering along axis b and to a significant increase in the background of low-temperature value of $\varepsilon'$. For ECMO, the free dispersion low-temperature values of $\varepsilon' \sim 100$–130 are higher than those values in EMO by a factors of 5‑6 in the dependence on the orientation of the crystal axes. [1]. Significant increase in dielectric permittivities $\varepsilon'$ at various frequencies along all the axes begins from $T \sim 175$ K. Near the room temperature, the maximum value of $\varepsilon'$ is observed along axis *b* ($3.5 \cdot 10^4$); $\varepsilon' \approx 1.75 \cdot 10^4$ along axes *a* and *c*.

The conductivities and their anisotropies in ECMO differ from those in EMO [1]. As well as in EMO, we are dealing with the real part of the conductivity $\sigma_1 = \omega\varepsilon''\varepsilon_0$ [30] that is calculated from dielectric losses $\varepsilon''$ (we measure the dielectric loss tangent $\tan\delta = \varepsilon''/\varepsilon'$). Here, $\omega$ is the circular frequency and $\varepsilon_0$ is the dielectric permittivity of free space. Conductivity $\sigma_1$ denoted below $\sigma$ is dependent on both the frequency and temperature. The conductivity of ECMO has the frequency dispersion similar to that observed in all $RMn_2O_5$ that we studied before [1‑4] (the higher the frequency the higher the conductivity). According to the accepted conductivity dispersion of inhomogeneous media containing local domains in a homogeneous matrix [30], the free dispersion low-frequency conductivity characterizes percolation conductivity of the matrix $\sigma_{DC}$. The $\sigma_{AC}$ conductivity depending on frequency (the higher the frequency, the higher the conductivity) belongs to the local conductivity of restricted domains. It is convenient to characterize both the ratio of the local and percolation conductivities and the temperature range of the existence of domains with local conductivity by the relative local conductivity $\sigma_{loc} = (\sigma_{AC} - \sigma_{DC})/\sigma_{DC}$ [30]. The temperature-frequency dependences of the relative local conductivities for various axes of ECMO are shown in the insets in Figs. 1b, 1d, and 1f.

The qualitatively similar dependences for the conductivities of ECMO are observed along axes a (Fig. 1b) and c (Fig. 1f). From Figs. 1b, 1f, it is seen that the two temperature ranges exist, in which the character of the frequency dependence of the conductivity is changed. At low temperatures up to some temperatures, which are dependent on the orientation of the crystal axes, the conductivity decreases as the frequency increases; i.e., at these temperatures, $\sigma_{DC} > \sigma_{AC}$. As temperature increases, the values of $\sigma_{AC}$ increase and, at some temperatures depending on the frequency, intersect the low-frequency conductivity $\sigma_{DC}$. The higher the frequency, the higher the temperatures at which such intersections are observed. The temperatures at which $\sigma_{AC}$ become equal to $\sigma_{DC}$ obey the Arrhenius law $\omega = \omega_0 \cdot \exp(-E_A/kT_m)$ (where $\omega$ is the frequency and $T_m$ is the temperature at which $\sigma_{AC} = \sigma_{DC}$, and $E_A$ is the activation energy at the boundaries of the local domains). This fact makes it possible to determine the activation barriers at the boundaries of these low-temperature local domains for crystal axes a and c (the insets in Figs. 1b and 1f). As is seen, there are additional high-temperature maxima of $\sigma_{loc}$, and the shifts of the temperatures of these maxima in the dependence on frequency also obey the Arrhenius law, which allows us to determine the activation barriers at the boundaries of these high-temperature local domains.

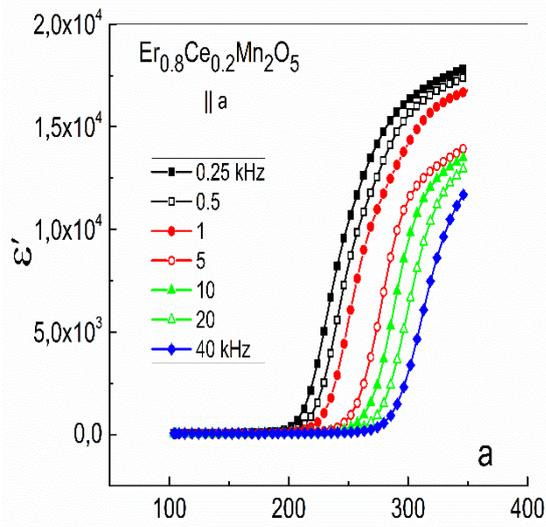
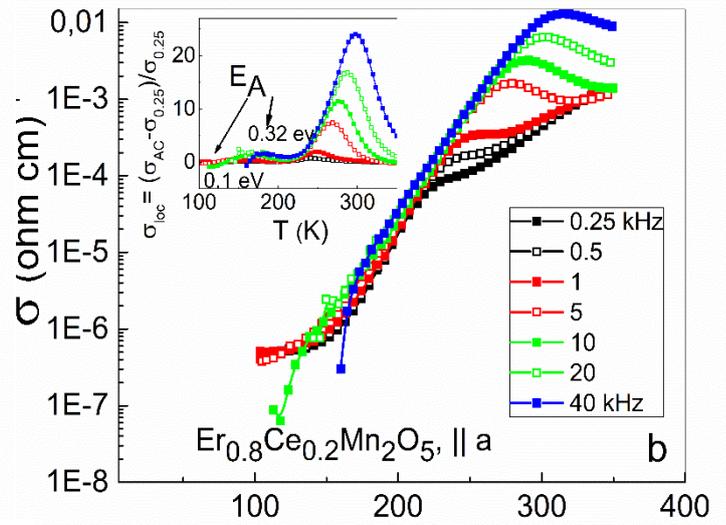
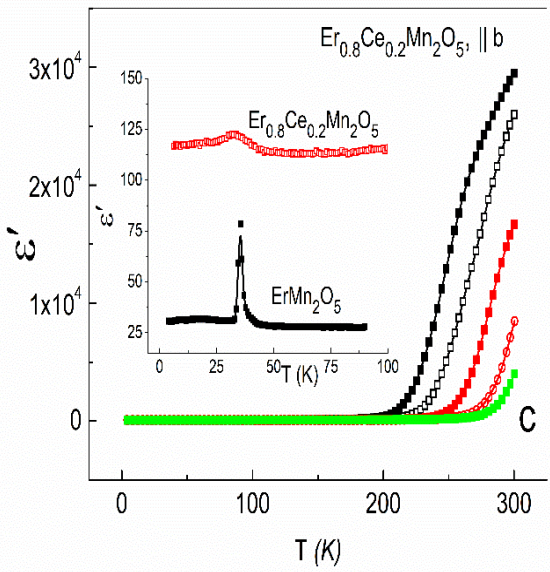
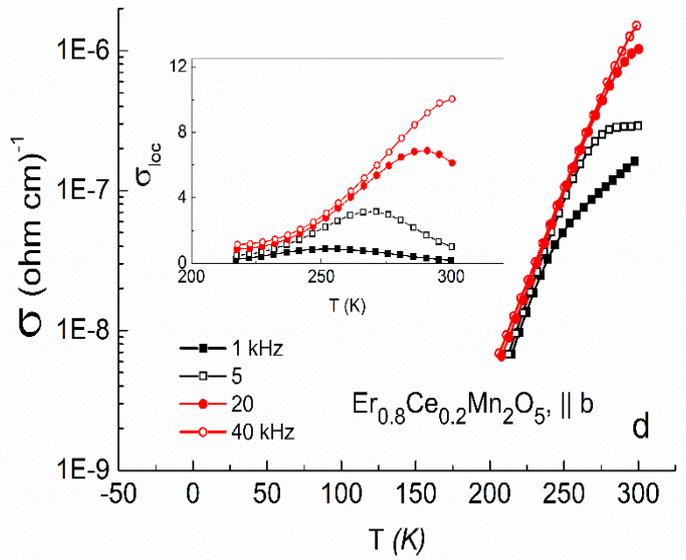
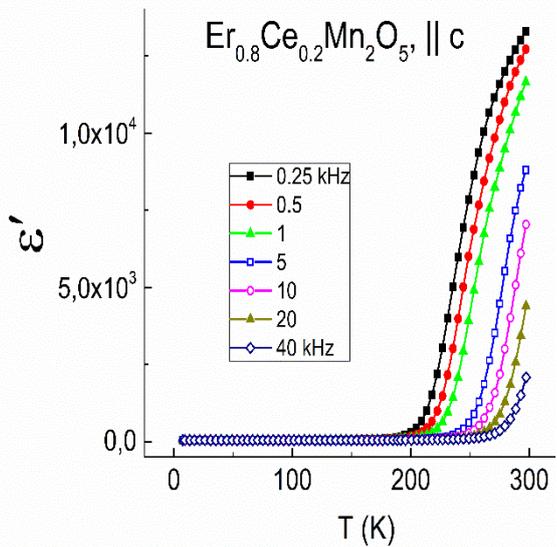
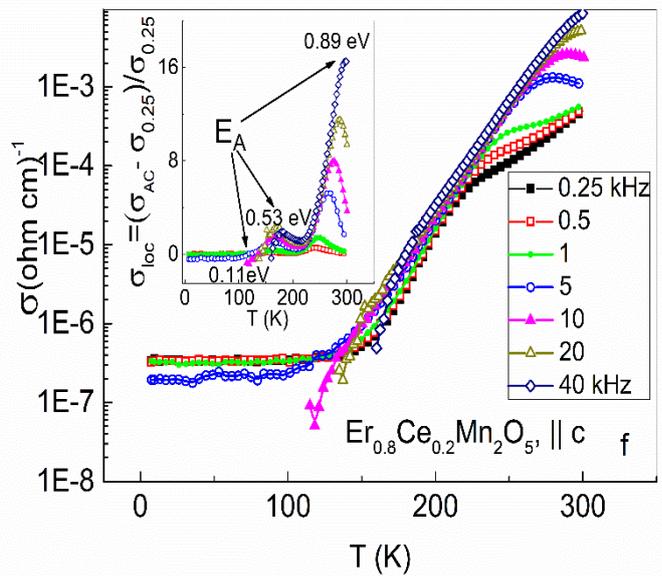

**Fig. 1.** Temperature dependences of dielectric permittivity ε' of $Er_{0.8}Ce_{0.2}Mn_2O_5$ for a number of frequencies along axes a, b, and c (1a, 1c, and 1e, respectively) and conductivity σ along the same axes (1b, 1d, and 1f, respectively). The inset in 1c demonstrates, in an enlarged scale, the temperature dependences of ε' for Er0.8Ce0.2Mn2O5 and ErMn2O5 in the temperature range 5–80 K at a frequency of 5 kHz. The insets in 1b, 1d, and 1f show the temperature dependences of the local conductivity. The frequencies are indicated in the plots.

They are also shown in the plots for local conductivities. Thus, the observation in ECMO of the local conductivity along axes a and c with different activation barriers demonstrates the existence of the phase separation domains of two types. We refer the domains with the negative local conductivity $σ_{loc}$ (i.e., with low-frequency percolation conductivity $σ_{DC}$ that is higher than high-frequency conductivity $σ_{AC}$) to the local phase separation domains that are formed inside the manganese subsystem that exists at low temperatures under high lattice barriers due to Er and Ce ions. As it was shown in [2, 3, 11, 12, 14–16] for $RMn_2O_5$ (R = Eu, Gd, and Bi), at low temperatures, the phase separation domain occupy a small volume of the crystal. The probability that the charge carriers will overcome the higher-temperature barriers in the lattice due to Er and Ce ions increases with temperature. In this case, at higher temperature the local phase separation domains are formed (Figs. 1b, 1f). Note that the conductivity in these domains also is due to the transport of $e_g$ electrons between manganese ions of various valences, but already in the domains with higher barriers. Thus, the local phase separation domains of the Mn subsystem exhibit different properties at fairly low and at higher temperatures after overcoming the lattice barrier. The main characteristics of these domains are the temperature dependences of local conductivities $σ_{loc}$ (the insets in Figs. 1b and 1f) and the electric polarization (Figs. 2a and 2b).

Figures 2a and 2b show, respectively, the temperature dependences of the remanent polarization of the loops of electric polarization Prem along all axes of ECMO and, as an example, the electric polarization loops for a number of characteristic temperatures along axis b. Note that the hysteresis loop shapes along all axes are similar, and the polarization values can be judged on the temperature dependence of the remanent polarization (Fig. 2a).

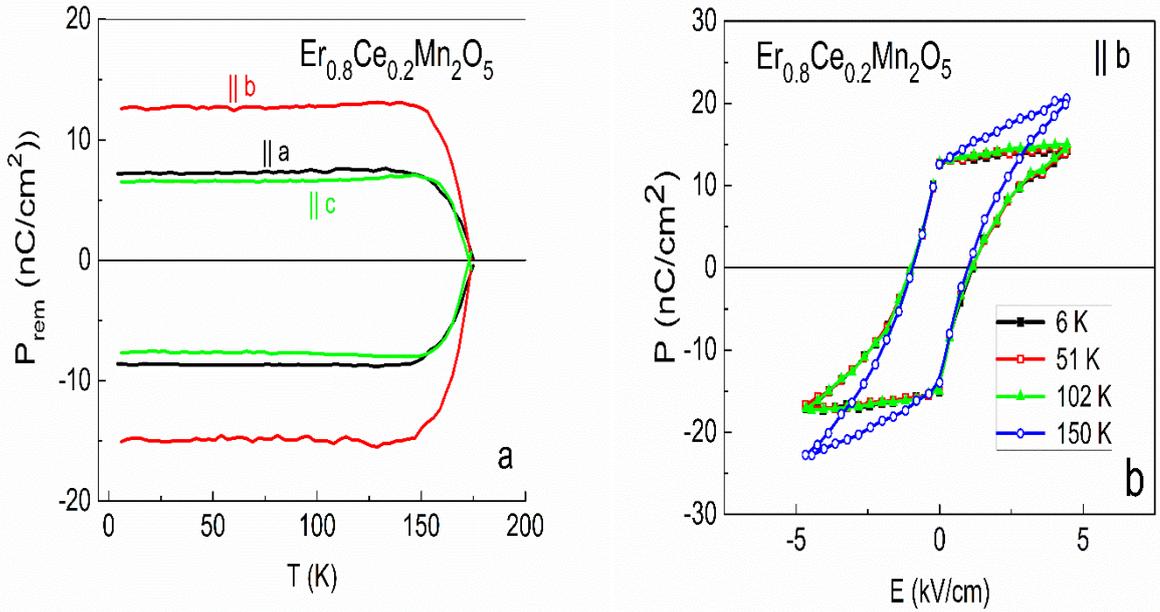

**Fig.2.** Temperature dependences of remanent polarization Prem of $Er_{0.8}Ce_{0.2}Mn_2O_5$ along various axis (2a) and a set of the hysteresis loops of the electric polarization along axis b at various temperatures (2b).

The conductivities along axes a and c (Figs. 1b and 1f) are almost the same. The activation barriers at the local domain boundaries at temperatures lower than 200 K (they are determined by the shifts of the first maxima of the relative local conductivity with frequency) are quite high (0.32 eV along axis a and 0.53 eV along axis c). The values of $\sigma_{loc}$ in the maxima near 175 K are 1.8 along axis a and 2.3 along axis c, what determines proximity of residual polarizations: 7.3 nC/cm$^2$ along the a axis and 6.5 nC/cm$^2$ along the c axis (Fig. 2a). The high barriers at the boundaries of the low-temperature local domains along axes a and c at an increased conductivity at all frequencies (Figs. 1b, 1f) provide the appearance of a relatively high carrier concentration under these barriers. The minimum remanent polarization along axis c in ECMO (Fig. 2a) at a higher barrier at the boundaries shows that the electron concentration under these barriers is most likely higher than the number of $Mn^{3+} - Mn^{4+}$ ionic pairs, which are recharged with electrons. These excess electrons screen, in partly, the internal electric fields inside local domains and decrease the polarization.

On the other hand, an anomalously low conductivity is observed in ECMO along axis b, and we cannot measure at T < 200 K both σ and $\sigma_{loc}$, as well as we cannot determine the barriers at the local domain boundaries. However, the existence of such domains evidenced by the character of the frequency dispersion of conductivity in Fig. 1d that is characteristic of local domains (the higher the frequency, the higher the conductivity) at T > 200 K (in common superparaelectric state). On the other hand, in the frozen superparaelectric state (at T < 175 K), the remanent polarization along axis b in ECMO is maximum (Prem ≈ 12.6 nC/cm$^2$) (Fig. 2a), while it was minimum along axis b in EMO (Prem ≈ 2 nC/cm$^2$), but it existed to higher temperature 270 K [1]. We believe that the lattice of ECMO is distorted more significantly in the direction of axis b as compared to that in EMO [1]. This fact should be referred as due to the influence of Ce ions, since Er ions maximally distort the lattice along axis *c*, and the maximum polarization 7.5 nC/cm$^2$ is precisely observed in EMO along axis *c* [1]. In this case, the main contribution to the polarization along axis b in ECMO is due to lattice distortions in the local domains from Ce ions, and the internal field in these domains and the activation barriers at their boundaries must be maximal. In this case, supposedly, a high charge carrier concentration must be accumulated inside the local domains and, correspondingly, the polarization screening must be increased. But the screening of the internal fields in these domains is insignificant, because of anomalously low conductivity along axis *b* (Fig. 1d)

In [1-4], it was shown for RMn$_2$O$_5$ (R = Gd, Bi, Er, and Tb) that the frozen superparaelectric state, for which the electric polarization loops were observed, exists up to the temperature, at which condition kT ≈ E$_A$ is fulfilled (E$_A$ is the activation barrier at the polar domain boundaries). At higher temperatures, the phase separation domains of the Mn subsystem continue to exist being in the usual superparaelectric state, for which the electric polarization disappears. This result agrees with theoretical work [17].

The increase of the dielectric permittivity of ECMO above the stable low-temperature background value also begins near 175 K. We focus attention to the fact that the anisotropy of the observed values of dielectric permittivity ε' at T > 175 K does not correspond to the observed anisotropy of the conductivities. This fact also shows that the values ε' are mainly determined by the structural distortions in the lattice, which are most strong along axis b.

Thus, along axes a and c in ECMO, there is a clear correlation of the dielectric properties and the conductivity of the local polar phase separation domains with the electric polarization induced by these domains. The same correlation was observed in $RMn_2O_5$ (R = Gd, Bi, Er, and Tb) studied before along all crystal axes [1‑4].

A qualitatively different situation appears in ECMO along axis b. As follows from the inset in Fig. 1c, the dilution of EMO with Ce ions strongly degrades the low-temperature ferroelectric transition. This fact demonstrates the distortion of the charge ordering of alternating ferromagnetic and antiferromagnetic $Mn^{3+}$‑$Mn^{4+}$ ion pairs along this axis. In EMO (as well as in any $RMn_2O_5$) such ordering led to the exchange-striction mechanism of the low-temperature ferroelectric ordering. It is the high probability of the $e_g$ electrons transport between the neighboring $Mn^{3+}$‑$Mn^{4+}$ ion pairs that leads to an increased background of the percolation conductivity along axis b in EMO, which decreases the local conductivity and the remanent polarization [1]. According to Fig. 1d, the conductivities of ECMO along axis b are two orders of values lower at all frequencies at temperatures above 200 K. This fact demonstrates either a significant fracture of the charge ordering along axis b or the appearance of quite strong distortion of the crystal field along axis b by Ce ions. The observation of a constant remanent polarization in ECMO along axis b at low temperatures up to 125 K (Fig. 2a) shows the existence of similar local polar phase separation domains from the lowest temperatures, but under very high lattice barriers formed by Ce ions. Thus, the distortion of the crystal field along this axis by Ce ions is the predominant factor along this axis.

As mentioned above, the Ce ions in our crystals have the variable valence $Ce^{+3.75}$ (i.e., a large part of Ce ions manifest themselves as $Ce^{4+}$ and partially as $Ce^{3+}$). Now we consider the sources of the distortion of the crystal field by Ce ions. Recall that R-ions in $RMn_2O_5$ have non-central pentagonal oxygen environment similar to that of $Mn^{3+}$ ions. In this case, they are disposed in layers with z = 0c. The $Mn^{4+}$ ions with the octahedral oxygen environments are contained in the neighboring planes with z = 0.25c and z = 0.75c. As a result of substitution of $Ce^{4+}$ ion for $Er^{3+}$ ion, an excess electron appears and it recharges ion $Mn^{4+} + e = Mn^{3+}$, which is the reason of the increase in the concentration of the local polar phase separation domains in ECMO as compared to that in EMO. We note, as well, that in spite of the close values of the $Ce^{4+}$ and $Er^{3+}$ ion radii (in the hexagonal configuration, they are 0.87 Å and 0.89 Å, respectively [24]) the crystal field in ECMO is distorted additionally as compared to that in EMO, for two reasons. First, this is due to an increase in the concentration of polar domains and, second, to the fact that larger $Ce^{3+}$ ions (with ion radius 1.01 Å) [24]) are substituted in partly for $Er^{3+}$ ions, although with a lower probability. These ions contain lone pairs of $6s^2$ electrons at the outer shell. The ions containing lone pairs of $6s^2$ electrons at the outer shells are known to strongly locally distort the lattice [22]. Thus, the concentration of the local polar phase separation domains in the ECMO solid solution is higher as compared to that in the initial EMO, and they are located in another distorted crystal field, which changes the dielectric properties and the electric polarization in ECMO as compared to EMO. The polarizations of these two crystals differ in the value, anisotropy, and temperature of existence. As it turned out, the $Er^{3+}$ ions distort the crystal field along axis c, and the $Ce^{+3.75}$ ions along axes a and b. The Mn ion subsystem disposed in the crystal fields formed by R ions is responsible for the formation of the local phase separation domains. Two of the following factors are sources of the polarity of the manganese systems themselves. Inside the phase separation domains, the double exchange related to the $e_g$ electron transport between the $Mn^{3+}$-$Mn^{4+}$ ion pairs leads to the fact that the positions of $Mn^{4+}$ ions (oxygen octahedra) are occupied by the Jahn‑Teller $Mn^{3+}$ ions which deform these octahedra. In turn, the smaller (as compared to $Mn^{3+}$ ions) $Mn^{4+}$ ions are located in noncentral pentagonal pyramids and also deform them additionally [1-4].

## 3.2. Dielectric Properties and Electric Polarization in $Tb_{0.8}Ce_{0.2}Mn_2O_5$.

Figures 3a‑3f show the temperature dependences of dielectric permittivity $\varepsilon'$ for $Tb_{0.8}Ce_{0.2}Mn_2O_5$ (TCMO) for a number of frequencies along axes a, b, and c (Figs. 3a, 3c, and 3e, respectively) and conductivity $\sigma$ along the same axes (Figs. 3b, 3d, and 3f, respectively). The inset in Fig. 3c shows the values of $\varepsilon'$ in a larger scale for some frequencies in TCMO along axis *b*. As seen, the dilution of Tb ions with Ce ions leads to almost complete fracture of the low-temperature ferroelectric ordering along axis b (compare to Fig. 4c [1]). Dielectric permittivities $\varepsilon'$ begin to significantly increase also from T ~ 175 K at various frequencies and along all the axes as was the case in ECMO. Near room temperature, the maximum value of $\varepsilon'$ is observed along axis c ($4 \cdot 10^4$), and the minimum value, along axis b ($6 \cdot 10^3$).

As seen from Figs. 3b, 3d, and 3f, the conductivities and their anisotropy in TCMO differ from these values in $TbMn_2O_5$ (TMO) [1]. The temperature–frequency dependences of the conductivity in TCMO are qualitatively similar along all the axes. However, the values of the conductivities are substantially different: they are close along axes a and b, while the conductivity along axis c is two orders of value lower.

Two types of the local domains are also observed in TCMO: low-temperature domains in the temperature range 5–175 K and high-temperature domains at T > 175 K. In the temperature range 175–225 K, the exponentially increasing (linear in a logarithmic scale) conductivities are observed along all the axes. The insets in Figs. 3b, 3d, and 3f show the temperature dependence of local conductivity $\sigma_{loc.}$ At temperatures lower 125 K, the local conductivity is negative

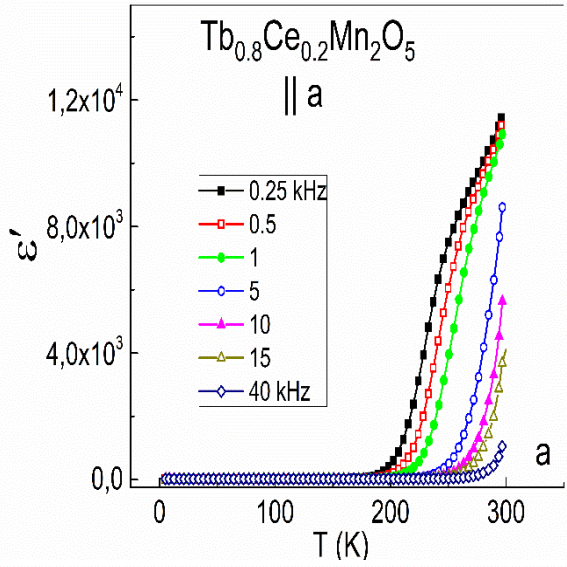
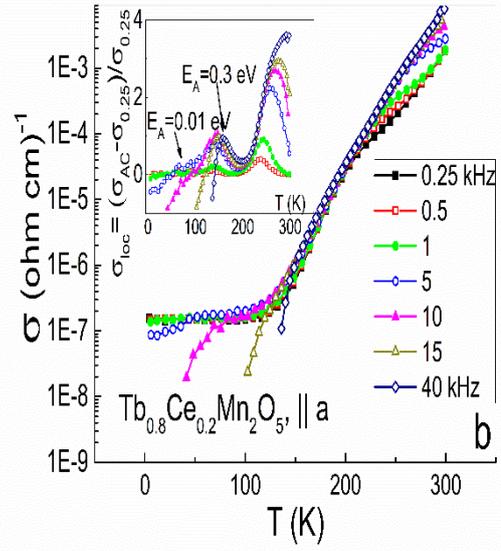
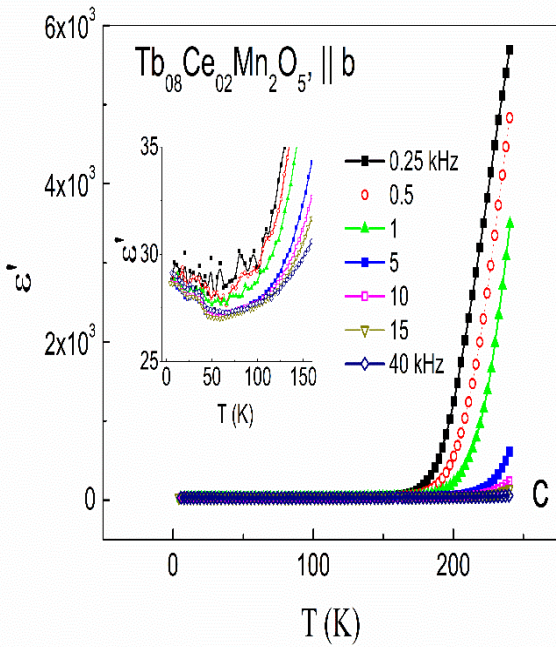
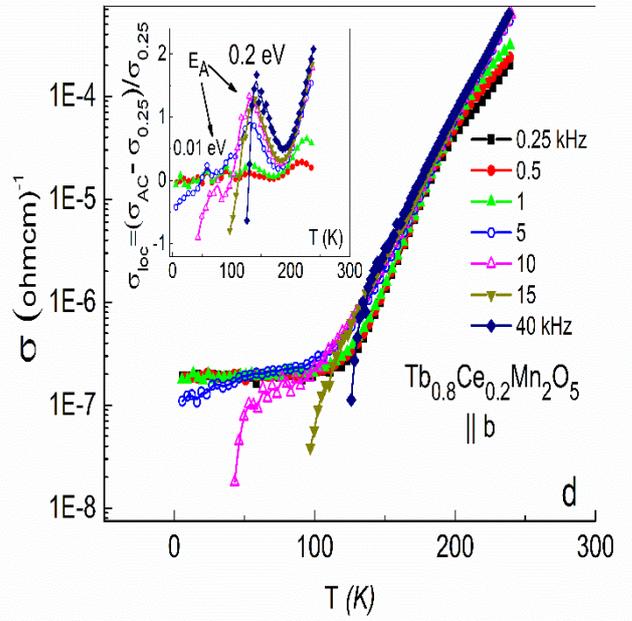
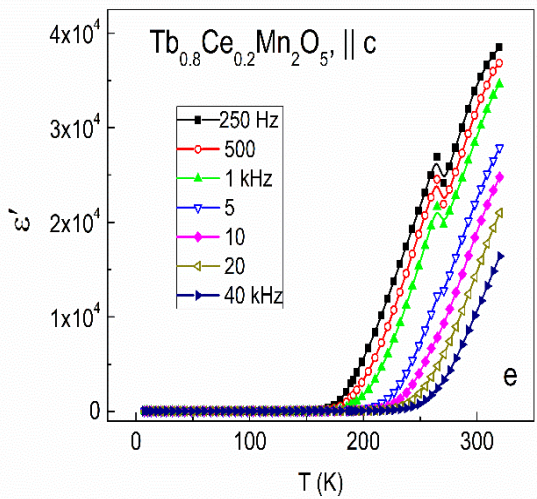
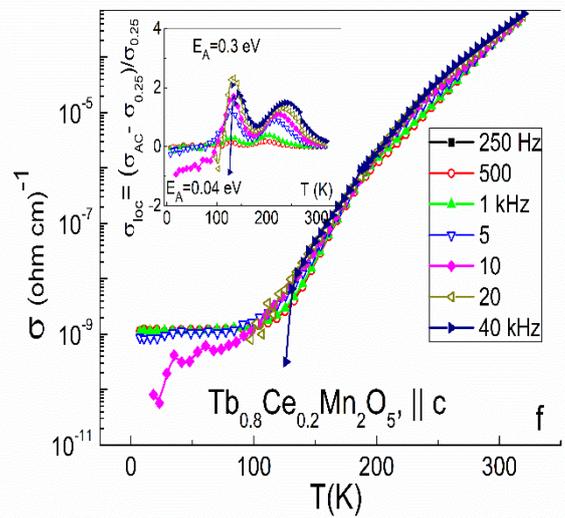

**Fig. 3.** Temperature dependences of dielectric permittivity ε' of Tb$_{0.8}$Ce$_{0.2}$Mn$_2$O$_5$ for a number of frequencies along axes a, b, and c (3a, 3c, and 3e, respectively) and conductivity σ along the same axes (3b, 3d, and 3f, respectively). The insets in 3b, 3d, and 3f show the temperature dependences of the local conductivity. The frequencies are indicated in the plots.

(i.e., the low-temperature percolation conductivity of the matrix is higher than the higher-frequency conductivities of the phase separation domains). Starting from the lowest temperatures, increasingly high-frequency conductivities at higher temperatures begin to cross the temperature-independent percolation conductivity, exceeding it. In this case, the local conductivity maxima are formed near 150 K. The high temperature slopes of the maxima approach the percolation conductivity as the temperature approaches T ~ 200 K (i.e., σ$_{loc}$ decreases approaching zero near 200 K). As mentioned above, the condition of the transformation of the frozen superparaelectric state to the common superparaelectric state and the disappearance of the remanent polarization is σ$_{loc}$ ≈ 0 as kT ≈ E$_A$ at the local domain boundaries.

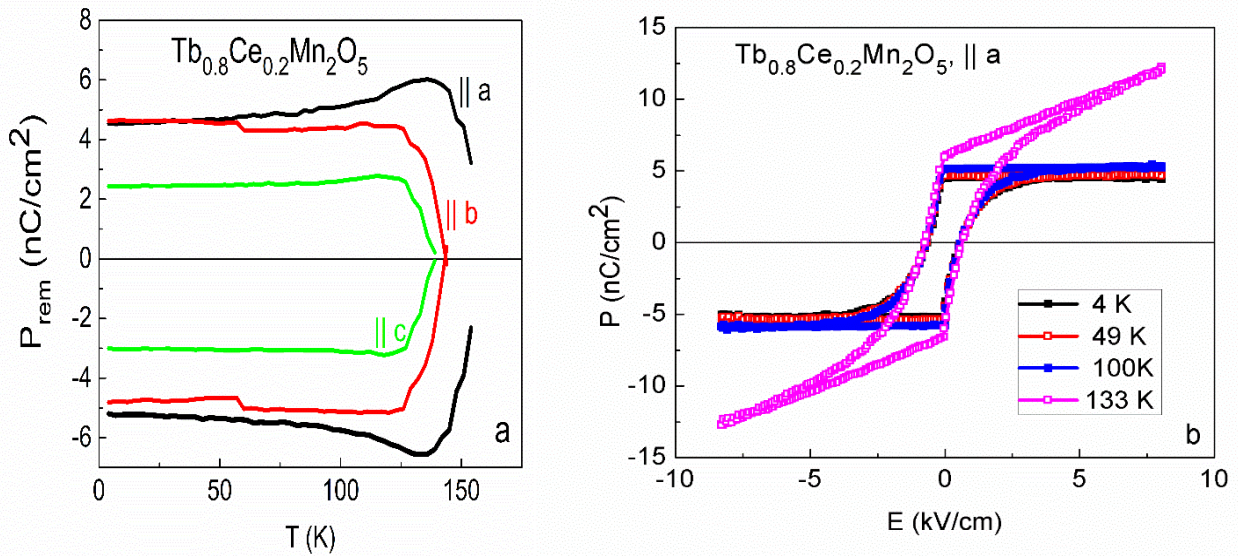

**Fig. 4.** Temperature dependences of remanent polarization *P*rem of Tb$_{0.8}$Ce$_{0.2}$Mn$_2$O$_5$ along various axis (3a) and a set of the hysteresis loops of the electric polarization along axis a at various temperatures (3b).

Figure 4a shows the temperature dependences of the remanent polarization along various axes in TCMO. Figure 4b shows the electric polarization loops in TCMO for some characteristic temperatures.

As seen from the comparison of Fig. 4a with the insets in Figs. 3b, 3d, and 3f, the temperatures of disappearance of the remanent polarization along various axes in TCMO are slightly lower than the temperatures, at which $\sigma_{loc}$ tends to a zero. They are close to 150 K. In this case, the maximum temperature is observed along axis a and the minimum one, along axis c. We consider the reason for this fact in more detail.

The conductivity along axis a is maximal, and the barrier at the boundaries of local domains responsible for the appearance of the remanent polarization is 0.3 eV (Fig. 3b and the inset in it). A close conductivity is observed along axis b, but the barrier at the boundaries is 0.2 eV (Fig. 3d). This fact provides similar background values of the low-temperature remanent polarization along axes a and b, but decreases the temperature of its existence along axis b (Fig. 4a). This result agrees with a decrease in activation barrier $E_A$ along axis b as compared to that along axis a.

The value conductivity along axis c is two orders lower than those along axes a and b (Fig. 3f), what decreases the concentration of the phase separation domains and therewith decreases the remanent polarization induced by these domains. However, both activation barriers at the boundaries of the local domains along the c axis, indicated in the inset in Fig. 3f, higher than along b axis: from $\sigma_{AC}$ intersections with $\sigma_{DC}$ is 4 times, from offsets the first maximum near 150K is 1.5 times. Seemed would, this should increase the temperature of existence polarization along the c axis compared to the b axis. On in fact, it decreases, albeit slightly (Fig. 4, a).

The reason for this is that the increase by four times of the low temperature barrier along the c axis leads to an increase in conductivity of the local domains under the high-temperature barrier (0.3 eV). Indeed, local conductivity in maxima (with an activation barrier of 0.3 eV) along axis c exceeds 2, while along axis a and b it is smaller (see inserts in Fig. 3, b, d, f). As a result excess electrons under the barrier 0.3 eV should him partially screen, lowering the actual temperature of the existence of residual polarization along the c axis (Fig. 4, a).

The ratios of both the activation barriers at the local domain boundaries in ECMO and TCMO are different significantly. These barriers in ECMO are significantly higher than they in TCMO. As a result, the concentration of carriers, that screen already low barriers, in TCMO increases. This reduces both the residual polarization values along all axes in the TCMO and the temperature of their existence. In ECMO, Er ions strongly distort the lattice along axis c, increasing the barriers at the domain boundaries along this axis. In the case of increased conductivity along axis $c$, this leads to screening the barriers and decreasing the polarization. At the same time, Ce ions in ECMO strongly distort the lattice along axis b, disturbing the charge ordering along this axis, sharply decreasing the conductivity. As a result, the maximum polarization in ECMO is observed along axis b.

In TCMO, Tb ions distort the lattice in plane ab at almost the same conductivities along axes a and b, and Ce ions strongly distort the lattice along axis c, which decreases the conductivity of the phase separation domains along this axis, which decreases the concentration of the phase separation domains and the polarization. As a result, the polarization in TCMO is significantly lower than that in ECMO. In this case, the maximum polarization is observed along axis a.

## 4. CONCLUSIONS

As follows from [1] and this work, the polar phase separation domains in $RMn_2O_5$ and $R_{0.8}Ce_{0.2}Mn_2O_5$ (R = Er, Tb) are formed in the Mn ion subsystem that consists of $Mn^{3+}$ and $Mn^{4+}$ ions distributed in the crystal in a certain way. The main process during formation of the polar phase separation domains is the electron transport between $Mn^{3+}$ and $Mn^{4+}$ ions that is substantially dependent on the values of the crystal fields induced by R and Ce ions. These local polar domains in the centrosymmetrical matrix of the crystals form the superparaelectric state that, below some temperatures, has the properties of the frozen superparaelectric state, whose response on an applied external electric field has the shape of the hysteresis loops of electric polarization with the remanent polarization. It is found that the polarization values and also the temperatures of their existence are determined by the character of crystal fields induced by R ions. In the case of dilution of Er and Tb ions with Ce ions the crystal field anisotropy in ECMO and TCMO are changed stronger than in EMO and TMO, and respectively, the values, the existence temperature of the electric polarization, and anisotropy also are more changed.

ACKNOWLEDGMENTS

This work was supported in part by the Russian Foundation for Basic Research (projects 18-32-00241) and program 1.4 of the Presidium of the Russian Academy of Sciences "Topical problems of the low-temperature physics."

CONFLICT OF INTEREST

The authors declare that they have no conflicts of interest.

REFERENCES

1. B. Kh. Khannanov, E. I. Golovenchits, and V. A. Sanina, Phys. Solid State **62**, 257 (2020).
2. B. Kh. Khannanov, V. A. Sanina, E. I. Golovenchits, and M. P. Scheglov, JETP Lett. **103**, 248 (2016).
3. B. Kh. Khannanov, V. A. Sanina, E. I. Golovenchits, and M. P. Scheglov, J. Magn. Magn. Mater. **421**, 326 (2017).
4. V. A. Sanina, B. Kh. Khannanov, E. I. Golovenchits, and M. P. Shcheglov, Phys. Solid State **60**, 537 (2018).
5. N. Hur, S. Park, P. A. Sharma, J. S. Ahn, S. Guba, and S.-W. Cheong, Nature (London, U.K.) **429**, 392 (2004).
6. Y. Noda, H. Kimura, M. Fukunaga, S. Kobayashi, I. Kagomiya, and K. Kohn, J. Phys.: Condens. Matter **20**, 434206 (2008).
7. P. G. Radaelli and L. C. Chapon, J. Phys.: Condens. Matter **20**, 434213 (2008).
8. P. G. de Gennes, Phys. Rev. **118**, 141 (1960).
9. L. P. Gor'kov, Phys. Usp. **41**, 581 (1998).

10. J. van den Brink and D. I. Khomskii, J. Phys.: Condens. Matter **20**, 434217 (2008).

11. V. A. Sanina, E. I. Golovenchits, V. G. Zalesskii, S. G. Lushnikov, M. P. Scheglov, S. N. Gvasaliya, A. Savvinov, R. S. Katiyar, H. Kawaji, and T. Atake, Phys. Rev. B **80**, 224401 (2009).

12. V. A. Sanina, E. I. Golovenchits, B. Kh. Khannanov, M. P. Scheglov, and V. G. Zalesskii, JETP Lett. **100**, 407 (2014).

13. M. Yu. Kagan and K. I. Kugel', Phys. Usp. **44**, 553 (2001).

14. E. I. Golovenchits, V. A. Sanina, and V. G. Zalesskii, JETP Lett. **95**, 386 (2012).

15. V. A. Sanina, E. I. Golovenchits, and V. G. Zalesskii, J. Phys.: Condens. Matter **24**, 346002 (2012).

16. V. A. Sanina, B. Kh. Khannanov, and E. I. Golovenchits, Phys. Solid State **59**, 1952 (2017).

17. M. D. Glinchuk, E. A. Eliseev, and A. N. Morozovska, Phys. Rev. B **78**, 134107 (2008).

18. V. Baledent, S. Chattopadhyay, P. Fertey, M. B. Lepetit, M. Greenblatt, B. Wanklyn, F. O. Saouma, J. I. Jang, and P. Foury-Leylekian, Phys. Rev. Lett. **114**, 117601 (2015).

19. N. Lee, C. Vecchini, Y. J. Choi, L. C. Chapon, A. Bombardi, P. G. Radaelli, and S. W. Cheong, Phys. Rev. Lett. **110**, 137203 (2013).

20. B. Kh. Khannanov, V. A. Sanina, and E. I. Golovenchits, J. Phys.: Conf. Ser. **572**, 012046 (2014).

21. K. P. Belov, A. K. Zvezdin, A. M. Kadomtseva, and R. Z. Levitin, *Orientational Transitions in Rare Earth Magnets* (Nauka, Moscow, 1979) [in Russian].

22. N. A. Hill and K. M. Rabe, Phys. Rev. B **59**, 8759 (1999).

23. Z. H. Sun, B. L. Cheng, S. Dai, K. J. Jin, Y. L. Zhou, Y. B. Lu, Z. H. Chen, and G. Z. Yang, J. Appl. Phys. **99**, 084105 (2006).

24. R. D. Shannon, Acta Crystallogr. A **32**, 751 (1976).

25. V. A. Sanina, L. M. Sapozhnikova, E. I. Golovenchits, and N. V. Morozov, Sov. Phys. Solid State **30**, 1736 (1988).

26. A. V. Babinskii, E. I. Golovenchits, N. V. Morozov, and L. M. Sapozhnikova, Sov. Phys. Solid State **34**, 58 (1992).

27. J. F. Scott, L. Kammerdiner, L. M. Parris, S. Traynor, V. Ottenbacher, A. Shawabkeh, and W. F. Oliver, J. Appl. Phys. **64**, 787 (1988).

28. M. Fukunaga and Y. Noda, J. Phys. Soc. J. **77**, 0647068 (2008).

29. S. M. Feng, Y. S. Chai, J. L. Zhu, N. Manivannan, Y. S. Oh, L. J. Wang, Y. S. Yang, C. Q. Jin, and Kee Hoon Kim, New J. Phys. **12**, 073006 (2010).

30. A. R. Long, Adv. Phys. **31**, 587 (1982).